\documentclass[letterpaper]{jpconf}
\usepackage{graphicx}

\begin{document}
\title{Infall, Fragmentation and Outflow in Sgr B2}
\author{Sheng-Li Qin$^{1,2}$, Jun-Hui Zhao$^{1}$, James M. Moran$^{1}$, Daniel Marrone$^{1}$,
N. Patel$^{1}$, Sheng-Yuan Liu$^{3}$,  Yi-Jehng Kuan$^{3,4}$, Jun-Jie Wang$^{2}$}
\address{$^1$ Harvard-Smithsonian Center for Astrophysics, 60 Garden St, Cambridge, MA 02138, USA}
\address{$^2$ National Astronomical Observatories, CAS, Beijing, 100012, P. R. China}
\address{$^3$ Academia Sinica Institute of Astronomy and Astrophysics, Taipei 106, Taiwan}
\address{$^4$ Department of Earth Sciences, National Taiwan Normal University, Taipei 116, Taiwan}
\ead{sqin@cfa.harvard.edu}
\begin{abstract}
Observations of H$_{2}$CO lines and continuum at 1.3 mm towards Sgr B2(N) and Sgr B2(M) cores were carried out with the SMA. We imaged  H$_{2}$CO line absorption against the continuum cores and the surrounding line emission clumps. The results show that the majority of the dense gas is falling into the major cores where massive stars have been formed. The filaments and clumps of the continuum and gas are detected outside of Sgr B2(N) and Sgr B2(M) cores. Both the spectra and moment analysis show the presence of outflows from Sgr B2(M) cores. The H$_{2}$CO gas in the red-shifted outflow of Sgr B2(M) appears to be excited by a non-LTE process which might be related to the shocks in the outflow. 
\end{abstract}
\section{Introduction}
The giant molecular cloud Sgr B2 is located in the Galactic center ($\sim$44$^{\prime}$ from SgrA*). High resolution radio continuum and recombination line observations at centimeter wavelenghts showed  the two most active star forming cores, Sgr B2(N) and (M), hosting clusters of massive stars associated with UCHII regions [1-6]. Previous observations showed that massive star formation is taking place in both Sgr B2(N) and (M), suggesting that the two hot cores are at different evolutionary stages [7-12]. H$_{2}$CO molecule has a simple chemical formation path and has been proven to be a useful probe of physical conditions of star formation regions [13,14]. We observed Sgr B2 at 1.3 mm with the SMA including the lines of H$_{2}$CO. The new results regarding the ongoing star formation in the Galactic center region are reported in this paper. 
\section{Observations and Results}
Two fields centered on Sgr B2(N) and (M)  were observed 
at 1.3 mm with the SMA\footnote {The Submillimeter Array is a joint project between the Smithsonian Astrophysical Observatory and the Academia Sinica Institute of Astronomy and Astrophysics and is funded by the Smithsonian Institution and the Academia Sinica.} in the compact configuration on August 1, 2005. 
Three H$_{2}$CO transitions were included in the 2 GHz sideband, one of 
them is blended with other molecular lines. The two unblended transitions 
are the H$_{2}$CO ($3_{03}-2_{02}$) and ($3_{21}-2_{20}$) with upper level 
energies (E$_{\rm u}$) of 21.0 K and 67.8 K at the rest frequencies 218.222 
and 218.760 GHz, respectively. The data reduction including 
calibration, imaging and analysis was carried out using {\it Miriad}. 
 The continuum subtraction was made in the UV domain using 
{\sc uvlin}. The mosaic of Sgr B2(N) and Sgr B2(M) 
was made using a simple linear mosaicing algorithm. The primary 
beam attenuation was also corrected. 

The mosaic of the continuum images of Sgr B2 with an angular 
resolution of 5$\rlap{.}^{\prime\prime}4\times 3\rlap{.}^{\prime\prime}2$
is shown in Fig. 1, showing the two bright components Sgr B2(N) 
(S$_{p}$=29.2$\pm$0.1 Jy~beam$^{-1}$) and Sgr B2(M) 
(S$_{p}$=20.2$\pm$0.1 Jy~beam$^{-1}$).
The continuum sources  
associated with the UCHII regions K1, K2 and K3
in the Sgr B2(N) core are not resolved at this angular resolution.
However a few continuum 
clumps have been detected, namely, 
Sgr B2(N)-E (S$_{p}$=1.1$\pm$0.1 Jy~beam$^{-1}$), Sgr B2(N)-W  
(S$_{p}$=1.0$\pm$0.1 Jy~beam$^{-1}$) and 
K4 (S$_{p}$=1.4$\pm$0.1 Jy~beam$^{-1}$). 
We have checked previous observations at longer wavelengths and
have found no radio continuum sources to be associated with Sgr B2(N)-E
and Sgr B2(N)-W, indicating that the continuum emission at 1.3 mm 
arises mainly from dust and the two clumps are probably in an 
early stage of star formation. 
The strong continuum emission from the unresolved 
Sgr B2(M) core mainly arises from the UCHII 
regions F1, F2, F3 and F4. In addition, a continuum component
(Sgr B2(M)-W, S$_{p}$=1.7$\pm$0.1 Jy~beam$^{-1}$) is 
located west of the Sgr B2(M) core. The position of the continuum peak at 1.3 mm of the Sgr B2(M)-W is consistent with that of the source B at 3.8 mm [9] and 
contains components A1 and A2 at 1.3 cm [2].
Finally, a component,
associated with the UCHII region Z10.24, is detected at
1.3 mm (S$_{p}$=2.0$\pm$0.1 Jy~beam$^{-1}$),
revealing a filamentary morphology located
between Sgr B2(N) and (M). 
\begin{figure}[h]
\includegraphics[width=16pc]{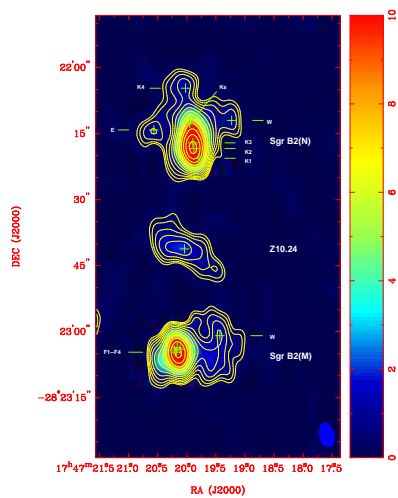}
\begin{minipage}[b]{16pc}
\caption{\label{label1} The mosaic of the Sgr B2 continuum image at 1.3 mm by combing 218 (LSB) and 228 GHz (USB) data. The synthesized beam is 5$\rlap{.}^{\prime\prime}4\times 3\rlap{.}^{\prime\prime}2$, PA=12.5$^{\circ}$ (lower-right corner). The wedge indicates the intensity scale from 0 to 10 Jy~beam$^{-1}$. The contours 
are -4, 4, 5.66, 8, 11.3, 16, 22.63, 32, 45.25, 64, 90.51, 128, 181.26 $\sigma$. The rms (1 $\sigma$) is 0.1 Jy~beam$^{-1}$.}
\end{minipage}

\end{figure}

The integrated H$_{2}$CO line emission flux images are shown in Fig. 2.  
Most of the emission is distributed around the two  cores with peak line 
emission flux of F$_{3_{03}-2_{02}}$=653 Jy~beam$^{-1}$~km s$^{-1}$ 
and F$_{3_{21}-2_{20}}$=696 Jy~beam$^{-1}$~km s$^{-1}$ for Sgr B2(N), 
and F$_{3_{03}-2_{02}}$ =92.6 Jy~beam$^{-1}$~km s$^{-1}$ and 
F$_{3_{21}-2_{20}}$=94.7 Jy~beam$^{-1}$~km s$^{-1}$ for Sgr B2(M). 
Clearly, the distribution of the H$_{2}$CO gas is not spherically 
symmetric with respect to each of the two star formation cores. 
In Sgr B2(N) region, in addition to the gas concentration on the core, 
there are a few clumps of gas distributed in the vicinity. We note 
that the emission images of the two transitions have a very similar 
morphology in the Sgr B2(N) region. In Sgr B2(M), the brightest 
emission of the higher transition gas is located 5$^{\prime\prime}$  
SE of the continuum peak while 
the peak emission 
of the lower transition gas coincides  with the continuum peak, 
suggesting that the two transitions in Sgr B2(M) correspond to 
different physical conditions. Most of the gas traced by low 
transition H$_{2}$CO line in Sgr B2(M) is located NW of the core, 
showing filamentary and clumpy structures. 
\begin{figure}[h]
\begin{minipage}[b]{100pc}
\includegraphics[width=32pc]{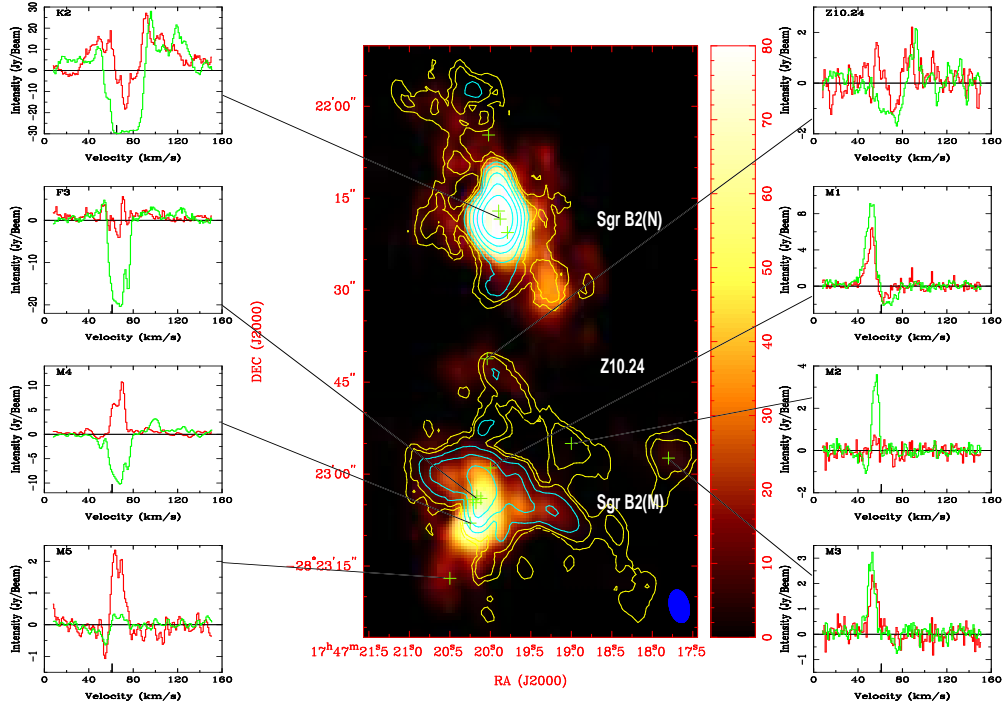}
\end{minipage}
\caption{\label{label2}Middle panel is the mosaic images of the
integrated intensities of H$_{2}$CO ($3_{21}-2_{20}$) (pseudo-color in a scale  
range from 0 to 80 Jy~beam$^{-1}$ km~s$^{-1}$) and 
 H$_{2}$CO ($3_{03}-2_{02}$) (contours 4,8,16,32,64,128,256 Jy~beam$^{-1}$ km~s$^{-1}$)
with FWHM beam of 5$\rlap{.}^{\prime\prime}4\times 3\rlap{.}^{\prime\prime}2$, PA=12$^{\circ}$.5. 
The left and right panels are the spectra at the positions of the various
components for the low transition $3_{03}-2_{02}$ (green) and  
the high transition $3_{21}-2_{20}$ (red), respectively. 
The horizontal line is the zero level after continuum-subtraction, 
and the vertical thick bars mark the systematic velocity of the sources. 
The systematic velocities are 65 and 61 km~s$^{-1}$ for Sgr B2(N) 
and Sgr B2(M), respectively.}
\end{figure}

The typical spectra at several positions are made from channel maps. 
The spectra towards the two continuum peaks  
in Sgr B2(N) and (M), show the gas components in either absorption 
or emission. The absorbing gas associated with two cores is dominated 
by the red-shifted gas with respect to the systematic velocity, 
suggesting that the gas is moving towards the cores. The spectrum of
the higher transition emission at K2 shows three absorbing peaks at
velocity 66.0, 73.0 and 80.4 km s$^{-1}$, respectively,
while the absorbing intensities of the lower transition appear to be 
saturated. The observed line intensity (F$_\nu$), the linewidth ($\Delta V$), 
the ratios of L/C and the ratio of F$_{3_{03}-2_{02}}$/F$_{3_{21}-2_{20}}$
give a strong constraint on the physical condition.
Considering the excitation condition and opacity effect of the two transitions,
the kinetic temperature of 40-50 K and H$_{2}$ density of 10$^7$ cm$^{-3}$ are suggested by a large-velocity-gradient (LVG) model for the absorbing gas in 
the core.
 The lower transition spectrum at F3 shows two absorbing peaks near 65 and 75 km s$^{-1}$ while the high transition spectrum show the absorption near 65 
km s$^{-1}$ and emission peak near 75 km s$^{-1}$. The multiple velocity 
components of the spectra in both absorption and emission appear to 
indicate that there are multiple gas clumps located at different depths 
inside the two cloud cores. High spatial resolution H$_{2}$CO line 
observations will be a critical key to resolve the gas kinematics of 
the multiple components in the two cores.    

The lower transition H$_{2}$CO ($3_{03}-2_{02}$) spectrum of Z10.24 shows 
that most of the gas is in absorption with a broad line width 
($\sim$50 km s$^{-1}$) and the emission is red-shifted with respect to 
the absorption, while the spectrum of the higher H$_{2}$CO transition 
shows a doubly peaked profile with an absorbing dip at 
65 km s$^{-1}$ which is consistent with optically thin 
radio recombination line H66$\alpha$ observations [5]. 
De Pree et al. (1996) [5] argued that an ionized outflow is
centered on Z10.24. The P-cygni profile in H$_{2}$CO ($3_{03}-2_{02}$) 
spectrum and doubly peaked profile in H$_{2}$CO ($3_{21}-2_{20}$) 
spectrum appear to favor the argument of an outflow from the 
UCHII region. The M1 located within the continuum core shows an inverse 
P-cygni profile in the spectra of both H$_{2}$CO transitions, 
providing evidence for the gas moving into the core. 
M2 located within the filament (30$^{\prime\prime}$ in length and 
7$^{\prime\prime}$ in width) shows significant emission 
(30 $\sigma$ in the peak line intensity) in the lower transition 
while the higher transition line emission is less significant 
(2-3 $\sigma$), suggesting that the gas is cold in this extended 
arclike structure. The line emission of both transitions from 
clumps M3 is detected ($>$ 15 $\sigma$). 

M4 and M5 reside in the outflow region of Sgr B2(M) [8, 11]. 
At M4 near the core, the spectra show that the high transition gas is 
in emission while the low transition gas is in absorption against 
the continuum. Most of the lower transition gas in absorption, 
which appears to be red-shifted with respect to the systematic velocity, 
is evident for the gas infalling into the core while the higher 
transition gas in red-shifted emission is suggested to be the 
outflow gas from the core. M5 is located at the tip of the outflow, 
where the spectrum shows a very significant line emission (20 $\sigma$) 
of the higher transition gas with a linewidth of $\sim$ 20 km s$^{-1}$, 
composed of  at least two velocity components while less significant 
emission (2-3 $\sigma$) of the lower transition gas is shown at 
the same spot, suggesting that the molecular gas does not satisfy the 
LTE condition. The H$_{2}$CO gas in the outflow appears to be excited 
by a non-LTE process, which might be related to the shocks in the outflow. 
The shocks could be C-shock if the gas is weakly ionized and shock 
velocity is below critical value $\sim$50 km s$^{-1}$ [15]. M5 is far away from the continuum at cm wavelengths and the H$_{2}$CO 
line width is below 20 km s$^{-1}$. The high transition H$_{2}$CO 
(3$_{21}$-2$_{20}$) in Sgr B2(M) outflow appears to be excited 
by a C-shock.  
 \begin{figure}[h]
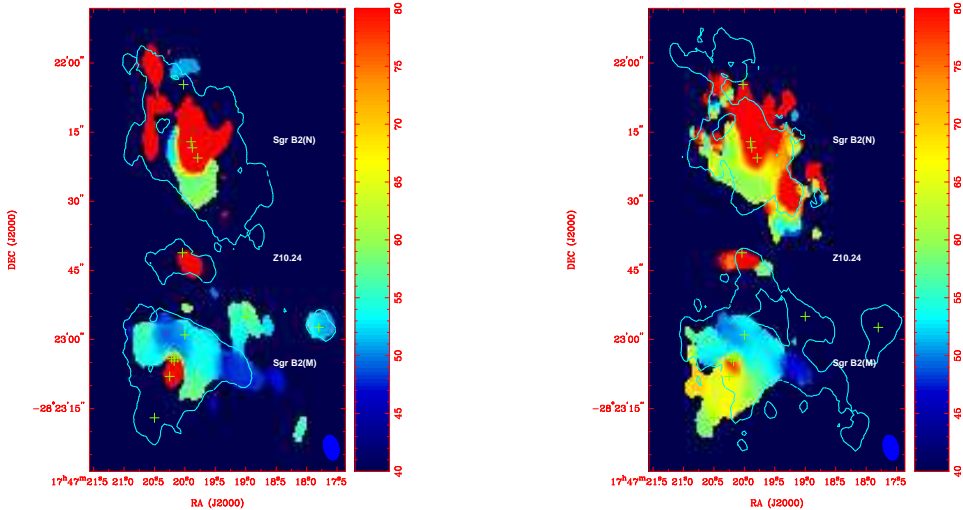

\begin{minipage}{18pc}
\centering
\includegraphics[width=16pc,angle=-90]{fig3a.eps}
\end{minipage}
\hspace{2pc}
\begin{minipage}{18pc}
\includegraphics[width=16pc,angle=-90]{fig3b.eps}
\end{minipage} 
\caption{\label{label1}The mosaic images of the 
intensity weighted velocity 
constructed using a cutoff of 8 $\sigma$
for the H$_{2}$CO ($3_{03}-2_{02}$) emission (color) overlaid with
the contours of H$_{2}$CO ($3_{21}-2_{20}$) intensity on the left panel.
The right panel is the velocity field of 
the H$_{2}$CO ($3_{21}-2_{20}$) emission  (color)
overlaid with the contours of H$_{2}$CO ($3_{03}-2_{02}$) intensity. 
The color wedge scales the 
velocity range from 40 to 80 km s$^{-1}$. 
The contours  outline the intensity level of 4 Jy beam$^{-1}$ km s$^{-1}$
for both images.}
\end{figure}
 
Fig. 3 shows the velocity field of H$_{2}$CO emission. The kinematical structure observed from the lower transition (see Fig. 3(a)) in Sgr B2(M) is relatively simple, consisting of a red spot 5$^{\prime\prime}$ SE of the compact core and the NW arch structure. The red spot (red-shifted emission, see also the spectrum M4 in Fig. 2) appears to be a fast moving compact component (V$_{lsr}\sim100$ km s$^{-1}$) ejected from the core. The appearance of the NW arch (blue-shifted emission) suggests that the gas is undergoing a complicated infalling process through interacting with the outflow while spiraling into the core rather than simply free falling. There is a velocity gradient in NS in Sgr B2(N) 
(see also Fig. 3(a)). Based on the HC$_{3}$N observations, Lis et al. 
(1993) [8] argued that the NS velocity gradient in Sgr B2(N) traces 
the rotation. The kinematics of the lower H$_{2}$CO transition in 
emission gas of Sgr B2(N) is likely dominated by the gas rotating around 
the core. The velocity field of the higher transition gas appears 
to trace the outflow better (see Fig. 3(b)). The red-shifted outflow SE 
to the core of Sgr B2(M) is clearly present. The kinematics of the NW 
arch of Sgr B2(M) observed in the high transition gas appears to 
be similar to that observed in the lower transition gas. The outflow 
in Sgr B2(N) was observed in SW direction [8]. 
The SE-NW velocity gradient of the high transition gas in 
Sgr B2(N) could be caused by the combination of the rotation and outflow.

Based on the line ratio of F$_{3_{03}-2_{02}}$/F$_{3_{21}-2_{20}}$ from the
spectra, most of the gas in the Sgr B2 region does not satisfy the LTE condition.
Assuming the molecular cloud is in plane-parallel geometry, the molecular 
excitation in non-LTE has been modelled based on an LVG approximation (e.g. 
[14],[16]). The non-LTE calculation suggests
 that the absorbing gas of the compact cores have a kinetic temperature of  $\sim$40 K and H$_{2}$ density of $\sim$10$^{7}$ cm$^{-3}$.
A higher kinetic temperature ($\geq$50 K) and relatively lower H$_{2}$ density ($\sim10^{6}$ cm$^{-3}$) are
 needed to explain
 the emission gas both residing in the cores (probably tracing
the outflows) and surrounding the cores. The detected larger-scale emission appears to be from 
a warm 
envelope, and absorbing gas appears to trace cold dense condensations 
associated with the massive star forming cores.
\section{Conclusions}
H$_{2}$CO lines and continuum at 1.3 mm were observed with SMA. The 
infalling gas was detected through the red-shifted absorbing gas against 
the continuum in Sgr B2 compact cores and their 
associated dusty clumps and filament Z10.24. The gas clumps 
in emission were also observed in both H$_{2}$CO lines. 
Based on an LVG model, our analysis
suggests that the gas in Sgr B2 complex consists of  cold dense cores and
a warm but less dense envelope. The red-shifted outflow in Sgr B2(M) is 
observed in H$_{2}$CO (3$_{21}$-2$_{20}$) line and appears to be 
excited by C-shock.  
\ack
We thank all the SMA staff of SAO and ASIAA for making SMA observations 
possible.

\end{document}